\documentclass[twocolumn,prl,showpacs,nofootinbib,amsmath,amssymb]{revtex4}

\usepackage{graphicx,epsfig,psfrag,axodraw,bm,amssymb}
\usepackage{dcolumn}
\usepackage{bm}
\usepackage{color}
\usepackage{mathrsfs,amsfonts,hepunits, color}

\long\def\comment#1{}

\begin{document}

\title{Dark Matter Detection with Polarized Detectors}

\author{Chi-Ting Chiang$^1$, Marc Kamionkowski$^{2,3}$, and
Gordan Z. Krnjaic$^2$}
\affiliation{$^1$Texas Cosmology Center and Department of
Astronomy, University of Texas at Austin, TX 78712}
\affiliation{$^2$Department of Physics and Astronomy, Johns
     Hopkins University, 3400 N.\ Charles St., Baltimore, MD
     21210}
\affiliation{$^3$California Institute of Technology, Mail Code
350-17, Pasadena, CA 91125}

\date{\today}

\begin{abstract}
We consider the prospects to use polarized dark-matter detectors to
discriminate between various dark-matter models.  If WIMPs are fermions
and participate in parity-violating interactions with ordinary
matter, then the recoil-direction and recoil-energy distributions
of nuclei in detectors will depend on the orientation of the
initial nuclear spin with respect to the velocity of the
detector through the Galactic halo. If, however, WIMPS are
scalars, the only possible polarization-dependent 
interactions  are extremely velocity-suppressed and, therefore, unobservable. 
 Since the amplitude of this polarization modulation is fixed by the detector speed through
the halo, in units of the speed of light, exposures several
times larger than those of current experiments will be required
to be probe this effect.
\end{abstract}

\pacs{98.80.-k}

\maketitle

Although dark matter has been known for several decades to
dominate the mass budget of galaxies, its particle nature is still
mysterious.  The coincidence between the interaction strength
required for an early-Universe relic to have the right cosmological
density and the electroweak interaction strength motivates the
idea that dark matter is composed of some weakly-interacting
massive particle (WIMP)
\cite{Jungman:1995df,Bergstrom:2000pn,Bertone:2004pz,DAmico:2009df}.  However,
WIMPs constitute a broad class
of dark-matter candidates, including heavy fourth-generation
neutrinos, various supersymmetric particles, particles in models
with universal extra dimensions \cite{ueds}, etc.; they may be scalar
particles or fermions, and if fermions, Majorana or Dirac
particles.  The precise nature of the couplings of dark matter to
ordinary particles varies considerably among the models.

An array of searches for WIMPs is now underway, but terrestrial
direct-detection experiments, designed to detect nuclear recoils
from collisions with dark-matter particles in the Galactic halo,  provide
likely our best hope to detect dark matter
\cite{Goodman:1984dc}.  These detectors
measure the energy of the nuclear recoils; the spectrum of such
recoil energies can then be used to discriminate a WIMP signal
from background, and in case of detection, to constrain WIMP
parameters and discriminate between different WIMP candidates.
It has also been suggested
\cite{Spergel:1987kx,Gondolo:2002np,Lisanti:2009vy} that the
{\it direction} of the nuclear recoil can additionally be used to
distinguish backgrounds and to constrain dark-matter
parameters, and this approach is now being implemented experimentally
\cite{Sciolla:2008rn}.

However, there is yet another handle these experiments can
exploit: the spin polarization of the detector nuclei. 
 If WIMPs
are scalar particles, then their interaction rate is essentially independent of 
the orientation of the nuclear spins. In scalar-nucleus scattering, 
the leading nuclear-polarization dependent terms arise from dimension 5 
operators and are
generically proportional to 
$| \vec{q}\,|^{2} (\vec q\cdot \vec s\,) \sim m_{N}^{3} v^{3} (\hat q \cdot \vec s \,)$ where $\vec q$ is the momentum transfer, $m_{N}$ is
the nuclear mass,  $v$ is the dark matter speed, 
and $\vec s$ is the nuclear polarization. Since this contribution to the total rate is ${\cal O}(v^{3}) \sim 10^{-9}$, its effects are
negligible in direct detection.   
However, if dark-matter particles are
fermions and if these particles have a
parity-violating interaction with ordinary matter, then the
total detection rate as well as the recoil and energy/direction
distribution can depend non-trivially on the polarization of the target
nuclei.  Thus, measuring the polarization dependence of these distributions
 may help discriminate between backgrounds and shed light on the WIMP's particle nature in
case of detection.
 
To illustrate, consider a toy model with a dark-matter particle $\chi$ of mass
$m_\chi$ that interacts with a spin $1/2$ nucleus $N$ of mass $m_N$ via the four-Fermi
operator,
\begin{equation}
      G \bar \chi \gamma^\mu (a+b\gamma_5) \chi \bar N
     \gamma_\mu (c+d \gamma_5) N~~,
\label{eq:lagrangian}
\end{equation}
where $G$ is a dimension $-2$ coupling constant, and $a$, $b$, $c$,
and $d$ are real parameters. For simplicity and without loss of 
generality we treat the nucleus as a 
point particle; the effects of all form factors and nuclear matrix elements are assumed to be
contained in the coefficients of our effective interaction. This interaction gives rise to a differential cross section,
\begin{equation}
    \frac{d\sigma}{dE}  = A + B\,(\!\vec{\, v} \cdot\! \vec{
    \,s} ) + B'( \vec v^{\,\prime}\cdot \vec s)+ {\cal O}(v^{2}),
 \label{eq:general}
\end{equation}
for a WIMP particle $\chi$ (antiparticle $\bar \chi$) of incident velocity $\vec v$ to scatter a nucleus,
initially at rest with polarization $\vec s$, to a recoil
energy $E$ and a final WIMP velocity $\vec v^{\,\prime}$.  Here,
\begin{eqnarray}
     A & =& \frac{G^2 m_N}{2\pi v^2} \biggl[ (a^2+b^2) (c^2+d^2) +
     (a^2-b^2) (c^2-d^2)  \nonumber \\
     &  & -  \frac{1}{2} (a^2+b^2) (c^2-d^2) -\frac{1}{2}
     (a^2-b^2) (c^2+d^2) \biggl], \nonumber \\
\end{eqnarray}
\begin{eqnarray}
     B_{\chi (\bar \chi)} & = & \frac{G^2 m_N}{2\pi v^2} \biggl[ cd(a^2+b^2) \pm
     ab(c^2+d^2)  \nonumber \\
     & &  \mp ab(c^{2}- d^{2})+ cd(a^2-b^2) \frac{m_\chi}{m_N} \biggr],
     \\
     {B'}_{\chi (\bar \chi)} & = & \frac{G^2 m_N}{2\pi v^2} \biggl[ cd(a^2+b^2) 
      \mp ab(c^2+d^2)  \nonumber \\
     & & \pm ab(c^{2}- d^{2})+ cd(a^2-b^2) \frac{m_\chi}{m_N} \biggr].
\end{eqnarray}
The dependence of the cross section, Eq.~(\ref{eq:general}), on
the dot product of a polar vector ($\vec v$ or $\vec v^{\, \prime}$) with an axial
vector ($\vec s\,$) is a manifestation of parity violation.  Parity 
violation requires that at least three of the parameters
 $a$, $b$, $c$, and $d$ be nonvanishing.  While we leave a discussion of detailed models
to future work, we do note that currently acceptable versions of
Dirac-neutrino dark matter
\cite{Belanger:2007dx,Schuster:2005ck} have such
couplings. Related parity-violating couplings may also be found in recent models of
 composite dark matter \cite{composite} and models with light force carriers in the dark sector \cite{Davoudiasl:2012qa}.
In the maximal-parity-violating case with a matter-antimatter asymmetry (no $\overline \chi$), we see
 that
 $a=-b=c=-d=1/2$, $A = B = G^{2} m_N (8\pi v^{2})^{-1}$ and $B'=0$.  We will assume this
case for our numerical work below.

\begin{figure}[t]
\SetScale{0.1}
\includegraphics[width=0.48\textwidth]{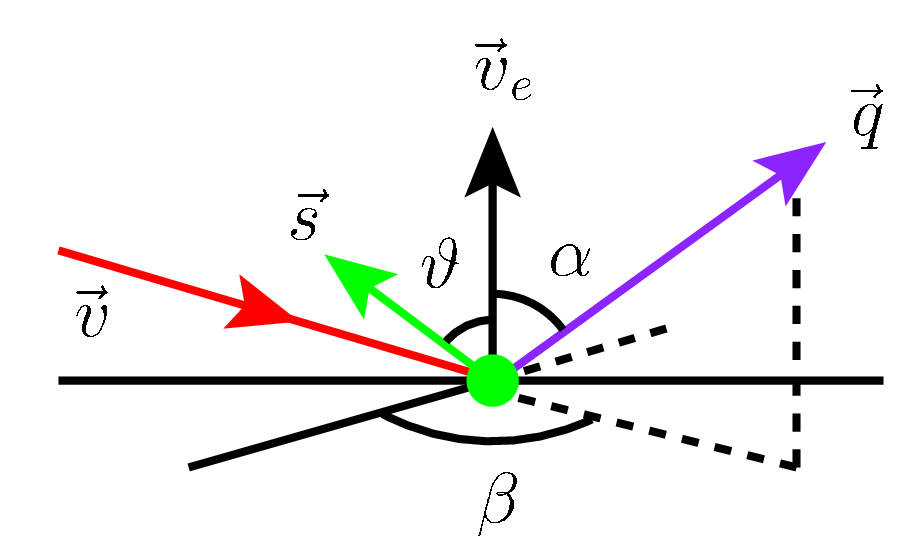}
\caption{An incident dark-matter particle with initial velocity $\vec v$
     scatters from a detector nucleus  with polarization $\vec s$
     and recoil momentum $\vec{\,q}$. Earth's velocity
     $\vec{\,v}_{e}$ is chosen to lie along the $z$-axis and the
     recoil rate in Eq.~(\ref{eq:mainrate}) is differential in
     both polar angle $\alpha$ and azimuthal angle $\beta$.}
\label{fig:scattering}
\end{figure}

\begin{figure}[t]
\includegraphics[width=0.5 \textwidth]{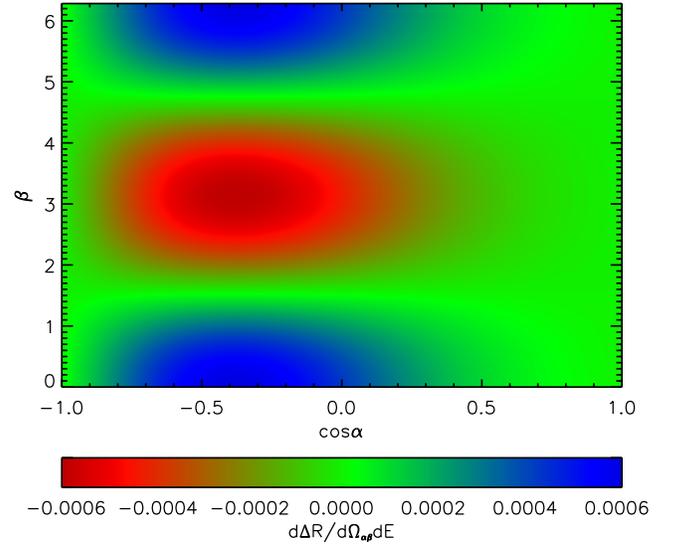}
\caption{Contour plot of the differential event rate,
     Eq.~(\protect\ref{eqn:finalresult}), in units of events/kg/day/keV/sr 
     as a function of $\cos\alpha$, where $\alpha$ is the polar
     angle, and the azimuthal angle $\beta$, for fixed recoil
     energy $E=30$~keV.  We take the angle between the
     polarization and the detector velocity through the Galactic
     halo to be $\vartheta=90^\circ$.  We also take
     $G=(100\,{\rm GeV})^{-2}$, and $m_\chi=100$~GeV.}
\label{fig:alpha-beta}
\end{figure}

We now calculate the distribution of recoil energies and
directions assuming that the detector moves with velocity $\vec
v_e$  (which throughout we will take to be
along the $\hat z$ axis) through the Galactic halo.
If we were to ignore directional
information, then we would simply calculate a (single)
differential event rate $dR/dE$.  If we were to consider the
combined recoil energy/direction distribution for an unpolarized
detector (as considered in
Refs.~\cite{Spergel:1987kx,Gondolo:2002np}), then we would
calculate a double-differential rate $dR/dE/d\cos\alpha$, where
$\cos\alpha\equiv \hat v_e \cdot \hat q$, and $\hat q$ is the
direction of the nuclear recoil.  If, however, the detector has
a spin polarization $\vec s$, which we take to be in the $x$-$z$
plane, at an angle $\vartheta$ from the $z$ axis, then there may
be an additional dependence on the
azimuthal angle $\beta$, about the $z$ axis, between
the recoil direction $\hat q$ and $\hat s$.  We must therefore
in this case calculate a {\it triple}-differential event rate
$dR/dE/d\cos\alpha/d\beta$.

We begin by writing the triple differential cross section $dR/dEd\Omega$ as
by demanding $\hat v \cdot \hat q =
q/2\mu v$, where $\mu$ is the reduced mass of the WIMP-nucleus
system, and $q$ is the recoil momentum.  Then \cite{Gondolo:2002np},
\begin{equation}
     \frac{d\sigma}{dE d\Omega} = \frac{1}{2\pi} \frac{d\sigma}{dE}
     \delta\left( \cos\gamma-\frac{q}{2\mu v} \right) = \frac{v}{2\pi}
     \frac{d\sigma}{dE} \delta(\vec v \cdot \hat q-v_q),
\label{eqn:doublecrosssection}
\end{equation}
where $v_q=q/2\mu$ is the nuclear recoil velocity, $d\Omega=
d\cos\alpha\, d\beta$ is a differential solid angle, and
$\delta(x)$ is the Dirac delta function.

The differential event rate \cite{Gondolo:2002np} per unit
detector mass is
\begin{equation}
     \frac{dR}{dE\, d\Omega } =\frac{n_{\chi}}{ 2 \pi m_N} \! \int
     \frac{d\sigma}{dE} \,\delta
     \left(  \vec v\cdot \hat q - v_{q} \right) v^{2} f(\vec v\,) \,
     d^{3}v \,,
\label{eq:mainrate}
\end{equation}
where $n_\chi$ is the local WIMP number density, and
$f(\vec v\,)$ is the dark-matter velocity distribution in the lab frame.
To isolate the polarization dependence in the general case, 
we subtract signals with opposite spin orientations,
\begin{eqnarray}
\label{eqn:eventrate}
     \frac{d\Delta R}{dE\, d\Omega } & \equiv&
     \frac{dR       (\vec s\,)     }{dE\, d\Omega } -  \frac{dR
     (-\vec s\,)   }{dE\, d\Omega} \\ \nonumber 
     & ~= &
      \frac{\, n_{\chi} }{  \pi m_N} \! \int  B\,(\vec v \cdot \vec s \,)\,\delta
      \left(  \vec v\cdot \hat q - v_{q} \right) v^{2} f(\vec
      v\,) \, d^{3}v.
\end{eqnarray}
We assume a standard Maxwellian halo, for which
\begin{eqnarray}
     f(\vec{v}\,) &=& \frac{1}{N}\,
     e^{-(\vec v+\vec v_{e})^{2}/ v_{0}^{2} }~~~~~ (|\vec v+\vec v_e| < v_{\rm esc}),  \\
     N&=& \pi v_{0}^{2}\left[ \sqrt{\pi} v_{0}\, {\rm
     Erf}(v_{\rm esc}/v_{0})  -2 \,   e^{-v_{\rm
     esc}^{2}/v_{0}^{2}}    \right],
\end{eqnarray}
where $v_{0} = 200$ km/s is the halo velocity
dispersion and $v_{\rm esc} = 500$ km/s is the escape speed from the Galactic halo.

To evaluate the integral in Eq.~(\ref{eqn:eventrate}), we choose
the spin vector to be in the direction $\hat
s=(\sin\vartheta, 0, \cos\vartheta)$, the initial WIMP-velocity 
direction to be $\hat v =( \sin\theta \cos \phi, \sin\theta
\sin\phi, \cos \theta)$, and the recoil direction to be  $\hat q =
(\sin\alpha \cos\beta, \sin\alpha \sin \beta, \cos\alpha)$. 
We then use the relation $\delta(g(v))=\delta(v-v_{1})/|g'(v_{1})|$ for the
Dirac delta function, where $g^{\prime}(v)$ denotes 
differentiation with respect to $v$, and $v_{1}$ satisfies $g(v_{1})=0$. This
occurs when $v =  v_q/g(\theta,\phi,\alpha,\beta) \equiv v_{1}$, where
$g(\theta,\phi,\alpha,\beta) = \cos\alpha\cos\theta+ \sin\alpha
\sin \theta \cos(\phi-\beta)$.  There is $\cos\theta$ dependence
in the integrand in Eq.~(\ref{eqn:eventrate}) through the
$\cos\theta$ dependence of $f(\vec v)$, and there is $\theta$
and $\phi$ dependence through 
\begin{equation}
     \vec v^{\, \prime} \cdot \vec s = v' s \left[\cos\theta \cos\vartheta +\sin\theta
     \sin \vartheta \cos\phi \right].
\end{equation}
Performing the $v$ integral in Eq.~(\ref{eqn:eventrate}), we
obtain
\begin{eqnarray}
     \frac{d\Delta R}{dE\, d\cos\alpha\, d\beta} &=&  \frac{G^2
     n_\chi}{8 N \pi^{2}
     }  \int_{-1}^1 dx \int_0^{2\pi} d\phi \frac{v'^2 \,
     \vec v^{\,\prime} \cdot \vec s}{ |
     g(\theta,\phi,\alpha, \beta)|} \nonumber \\
     &  &\times e^{-(v'^2+v_e^2+2v'v_e x)/v_0^2} \nonumber \\
     & & \times \Theta(v'-v_{\rm min}) \Theta(v_{\rm esc}-v'),
\label{eqn:finalresult}
\end{eqnarray}
where $v'$ and $g$ both carry dependence on $\theta$ and
$\phi$.  Here, $v_{\rm min}= (m_N E/2\mu^2)^{1/2}$ is the minimum
WIMP velocity\footnote{  The analysis can be generalized to inelastic dark
matter \protect\cite{TuckerSmith:2001hy} by using $v_{\rm min} = \left[
\Delta+ (m_N E/\mu) \right](2 m_N E)^{-1/2}$.} required to produce a nuclear recoil energy
$E$, and $\Theta(x)$ is the unit step function.
The polarization-induced breaking of the azimuthal symmetry
about $\vec v_e$ prevents us from going further analytically, as
can be done otherwise \cite{Gondolo:2002np}, but the remaining
double integral is straightforward to evaluate numerically.

\begin{figure}[t]
\includegraphics[width=0.5 \textwidth]{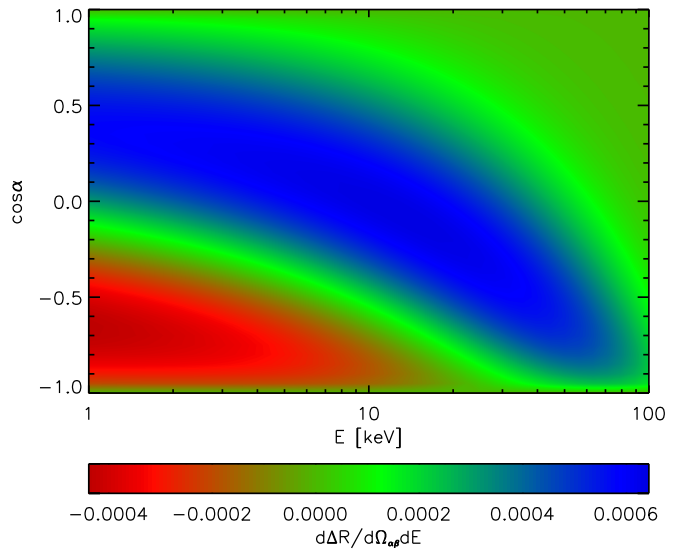}
\caption{Contour plot of the differential event rate,
     Eq.~(\protect\ref{eqn:finalresult}), in units of events/kg/day/keV/sr
     as a function of $\cos\alpha$ and the recoil energy $E$,
     for fixed azimuthal angle $\beta=0$.  We assume
     $\vartheta=90^\circ$, $G=(100\,{\rm GeV})^{-2}$, and
     $m_\chi=100$~GeV.}
\label{fig:alpha-E}
\end{figure}

\begin{figure}[t]
\includegraphics[width=0.5 \textwidth]{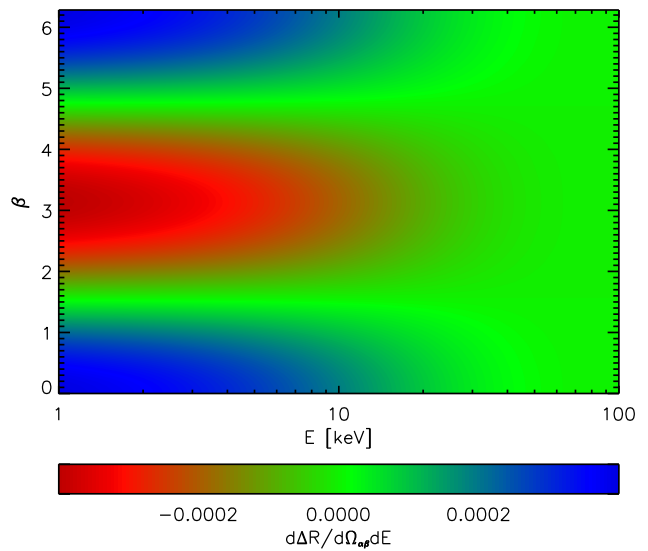}
\caption{Contour plot of the differential event rate,
     Eq.~(\protect\ref{eqn:finalresult}), in units of events/kg/day/keV/sr
     as a function of $E$ and $\beta$ for fixed polar angle
     $\alpha=45^\circ$.  We assume
     $\vartheta=90^\circ$, $G=(100\,{\rm GeV})^{-2}$, and
     $m_\chi=100$~GeV.}
\label{fig:E-beta}
\end{figure}

\begin{figure}[t]
\includegraphics[width=.48 \textwidth]{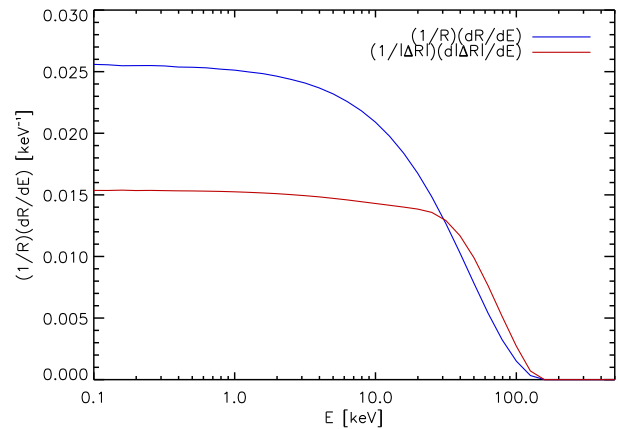}
\caption{ Plot of the differential event rate $(1/R)(dR/dE)$  
(red, color online) and polarization-dependent differential event rate
     $1/(|\Delta R|) d \!\left|\Delta R \right|/dE $ (blue) after numerically integrating over
      $\alpha$ and $\beta$. 
     As before, we assume  $\vartheta=90^\circ$, $G=(100\,{\rm GeV})^{-2}$, and
     $m_\chi=100$~GeV.  
     }
\label{fig:differentialE}
\end{figure}

Figs.~\ref{fig:alpha-beta}, \ref{fig:alpha-E}, and
\ref{fig:E-beta} show numerical results for $d\Delta
R/dE/d\cos\alpha/d\beta$, the polarization-dependent part of the
triple-differential event rate, as a function of the recoil
energy $E$, polar angle $\alpha$, and azimuthal angle $\beta$ of
the recoil nucleus.  The key feature here is the $\beta$
(azimuthal-angle) dependence in Figs.~\ref{fig:alpha-beta} and
\ref{fig:E-beta} which would not arise without parity-violating
interactions and particle-antiparticle asymmetry.  The
dependence of $dR/dE/d\cos\alpha$ also depends on the
polarization, as shown in Fig.~\ref{fig:alpha-E}, even without
azimuthal-angle information.  Fig. \ref{fig:differentialE}
shows the differential recoil spectra $(1/R)(dR/dE)$ (blue curve,
color online) and  $(1/ |\Delta R|) (d|\Delta R|/dE)$ (red
curve) obtained by integrating over the angular dependence in
Eqs. (\ref{eq:mainrate}) and (\ref{eqn:eventrate})
respectively. Thus, a polarization-dependent detection rate can
be sought even without directional information.

Since the detector will be fixed in the Earth frame, the
daily revolution of the Earth provides a natural modulation of
the detector orientation with respect to the Earth's velocity
through the halo.  This can be used to isolate the dark-matter
signal from systematic experimental effects (e.g., variations
in sensitivity of the detetor with recoil direction) that might
mimic such a signal.  The rotation of the Earth around the Sun
provides an additional systematic check.

A quick estimate shows that the experimental exposure needed 
to isolate polarization dependence may be feasible. Consider, for example, a 
WIMP candidate with total 
detection rate $R_{\rm tot}$. Since the polarization modulated
amplitude is velocity suppressed, $\Delta R/R_{\rm tot} \sim v
\sim 10^{-3}$, it is necessary to observe approximately $3
\times 10^{6}$ total events for a  $3\sigma$ discovery.  For the purpose of illustration, we can compare
this benchmark figure with those of the DAMA experiment.
If the DAMA NaI results are due to dark matter with a total
scattering rate $R_{\rm tot} \sim 0.5$ kg$\cdot$day with a
cumulative exposure of $\sim 1$ ton$\cdot$yr $\sim 4 \times
10^{5}$ kg$\cdot$day \cite{DAMA}, then a future experiment with polarized
nuclei observing the same dark matter interactions will
need to amass an exposure roughly an order of magnitude larger
than DAMA's to observe this signal.  While larger than those of
current detectors, this exposure is within  the scale of those
considered for development within the next decade
\cite{Baudis:2012bc}.  The detector would not
only have to be larger, but also be polarized and possibly have
direction sensitivity, neither of which are true of DAMA.  More
precise estimates of detection rates will also require
proper consideration of nuclear form factors and, depending on
the nucleus, of nuclear spins $J>1/2$.
 
While we have focussed here on WIMP-nucleon scattering, similar
ideas may apply to detection schemes based on
dark-matter--electron scattering \cite{Starkman:1995ye}, in which
the signal may conceivably be enhanced by the larger scattering
rates associated with more abundant lower-mass dark-matter
candidates.  The increased relative velocity between the
dark-matter particle and the target particle (an atomic
electron) may also enhance the polarization dependence.

\medskip
GK thanks Fabrizio Caola, David E.~Kaplan, and Kirill Melnikov
for helpful discussions.  GK is supported by the National
Science Foundation under grant number 106420.  MK is supported
by DoE DE-FG03-92-ER40701 and NASA NNX10AD04G. This research was 
supported in part by Perimeter Institute for Theoretical Physics, which is 
 is supported by the Government of Canada through Industry Canada
and by the Province of Ontario.

\end{document}